# Magnetic Resonance Imaging with a Dielectric Lens

F. Vazquez[1], O. Marrufo[1], R. Martin[1], and A. O. Rodriguez[1]
[1]Departament of Electrical Engineering, Universidad Autonoma Metropolitana Iztapalapa, Mexico, DF, Mexico

**Introduction.**

Recently, metamaterials have been introduced to improve the signal-to-noise ratio (SNR) of magnetic resonance images with very promising results ([1] and references therein). However, the use polymers in the generation of high quality images in magnetic resonance imaging has not been fully been investigated. These investigations explored the use of a dielectric periodical array as a lens to improve the image SNR generated with single surface coils. Commercial polycarbonate glazing sheets were used together with a circular coil to generate phantom images at 3 Tesla on a clinical MR imager.

**Material and Methods**.

Polycarbonate glazing sheets were used. These have a similar geometrical configuration to the perfect lens, since they both have a periodic structure (periodic chain of monomers). Lexan Thermoclear 2UV sheets (General Electric Structured Products, GE Co. Pittsfield, MA, USA) were used to act as a lens to concentrate the RF signals emanating from the sample and redirect them to the receiving-only surface coil. The sheets were 27x27cm with 0.5 mm thickness and filled with air. RF transmission was performed with a whole-body birdcage (68 cm long, diameter 66 cm and 16 rungs), and reception was performed with a circular-shape coil (12 cm diameter). Different configurations were tested: a) no glazing sheet, b) glazing and coil formed a parallel array where the phantom was located in-between them, and c) same array as in b), but rectangular cells were rotated $90^0$. All elements were properly fixed to avoid any movement between different experiments. Fig. 1 shows the experimental array used to acquire images. The phantom was 30 cm long and had a 12 cm diameter. All imaging experiments were carried out on a 3T commercial imager (Philips Medical Systems, Best, NL). T1-weighted images of a saline-solution phantom were acquired using spin echo sequences with the following acquisition parameters: TR/TE=450/10 ms, FOV=250x290, voxel size=0.9mm, slice thickness=4mm, NEX=1.

**Results and Discussion**.

T1-weighted phantom images were obtained for all three arrangements, as shown in Fig. 2. This image data was used to compute uniformity profiles to compare performance for the three cases, as depicted in Fig. 3. From these results, a significant improvement in the image SNR of the c)-configuration over the two others can clearly be seen. However, no improvement is observed for the b)-configuration; this is a rather intriguing result since we would have expected at least a small increase in SNR. Cell orientation seems to play an important role that should be further investigated. In addition, the size of cells and number of sheets should also be put to the test to study how these other two parameters affect the image SNR. We have experimentally demonstrated that the use of dielectric periodical arrays can produce a significant increase in the image SNR. This are rather interesting results since only a single surface coil was used, and also the glazing sheets were 12 cm from the coil. This seems to be working as a lens located at the limit of the penetration capability of the coil. Another advantage is that this glazing material is easily available. There is still much scope to test other possible configurations; for example, manufacturing cylinders made out of polycarbonate glazing to entirely cover the sample. Further investigation is required to explain the physical mechanisms involved in the improvement of the image SNR.

**Acknowledgments**.

F. V., O. M. and R. M. would like to thank CONACYT Mexico for Ph. D. scholarships. Email: arog@xanum.uam.mx.

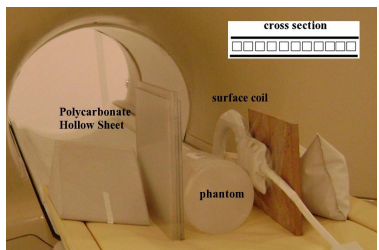

Figure 1. Experimental setup and a cross-section of glazing sheet.

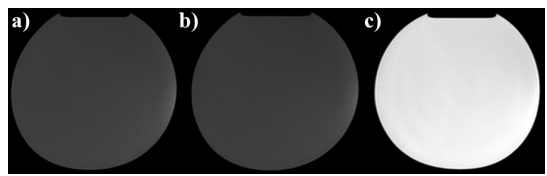

Figure 2. Phantom images of different cell positions: a) no panel, b) horizontal position, and c) cells rotated $90^0$ with respect to b) (see cross-section in Fig. 1).

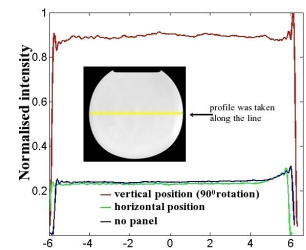

Figure 3. Comparison profiles for the positions in Fig. 2.